\documentclass[sigconf]{acmart}

\usepackage{booktabs} 
\usepackage{enumitem}
\usepackage{subfig}

\newcommand{\newcomment}[3]{%
  \expandafter\newcommand\csname #1\endcsname[1]{\textcolor{#2}{[#3: ##1]}}%
}
\newcomment{lavi}{red}{Lavi}
\newcomment{jevin}{teal}{Jevin}
\newcomment{kishore}{orange}{Kishore}

\setcopyright{rightsretained}

\acmConference[KDD: BigScholar 2018]{Special Interest Group on Knowledge Discovery and Data Mining's (SIGKDD's) BigScholar}{August 2018}{London, UK}
\acmYear{2018}
\copyrightyear{2018}

\begin{document}
\title[Is together better?]{Is together better? Examining scientific collaborations across multiple authors, institutions, and departments}

\author{Lovenoor Aulck}
\affiliation{%
  \institution{University of Washington Information School}
  \streetaddress{1851 NE Grant Lane}
  \city{Seattle}
  \state{Washington}
  \postcode{98105}
}
\email{laulck@uw.edu}

\author{Kishore Vasan}
\affiliation{%
  \institution{University of Washington Information School}
  \streetaddress{1851 NE Grant Lane}
  \city{Seattle}
  \state{Washington}
  \postcode{98105}
}
\email{kishorev@uw.edu}

\author{Jevin West}
\affiliation{%
  \institution{University of Washington Information School}
  \streetaddress{1851 NE Grant Lane}
  \city{Seattle}
  \state{Washington}
  \postcode{98105}
}
\email{jevinw@uw.edu}

\renewcommand{\shortauthors}{L. Aulck et al.}

\begin{abstract}
Collaborations are an integral part of scientific research and publishing. In the past, access to large-scale corpora has limited the ways in which questions about collaborations could be investigated. However, with improvements in data/metadata quality and access, it is possible to explore the idea of research collaboration in ways beyond the traditional definition of multiple authorship. In this paper, we examine scientific works through three different lenses of collaboration: across multiple authors, multiple institutions, and multiple departments. We believe this to be a first look at multiple departmental collaborations as we employ extensive data curation to disambiguate authors' departmental affiliations for nearly 70,000 scientific papers. We then compare citation metrics across the different definitions of collaboration and find that papers defined as being collaborative were more frequently cited than their non-collaborative counterparts, regardless of the definition of collaboration used. We also share preliminary results from examining the relationship between co-citation and co-authorship by analyzing the extent to which similar fields (as determined by co-citation) are collaborating on works (as determined by co-authorship). These preliminary results reveal trends of compartmentalization with respect to intra-institutional collaboration and show promise in being expanded.
\end{abstract}

%
%



\maketitle

\section{Introduction}

``Good teams'' are said to make for ``good science.'' But what makes a good team? What are the necessary ingredients and costs? There has long been interest in studying scientific collaborations \cite{beaver2001reflections, luukkonen1992understanding} as they have continually become the de facto way in which research is conducted \cite{o2012change}. In the past, access to large-scale corpora has limited the ways in which questions about collaborations could be investigated. However, with improvements in data/metadata quality and access, more refined questions can now be pursued. In this paper, we look at specific kinds of collaboration. Specifically, we look at \textit{multi-departmental} and \textit{multi-institutional} collaborations and ask whether the impact of these collaborations differs from those of general multi-author collaborations. In this paper, we fully disambiguate the department affiliations at a large public university in the United States (US), using a large-scale bibliographic database and local knowledge of the university. We then analyze the impact of multi-departmental and multi-institutional collaborations and how they compare to multi-author collaborations. We find that multi-departmental and multi-institutional collaborations follow similar patterns to other collaborations in terms of impact as collaborative papers tend to have more citations. 


Collaboration does not come free of costs as there are reasons to be wary of research collaboration, such as it potentially threatening the motivation and accountability of scientists \cite{wray2006scientific} and it slowing research endeavors \cite{fox1984independence}. However, there are also numerous perceived benefits to collaborative work. For example, research collaborations are believed to improve the quality and creative potential of scientific papers. With respect to quality, collaborative research tends to be cited more often than non-collaborative work \cite{beaver2004does, levitt2010does, fox1991gender, glanzel2001double, figg2006scientific} and citations are often assumed to correlate with quality \cite{beaver2004does}, rightly or not \cite{garfield1979citation}. The idea of associating collaborations with quality is also furthered by evidence that collaborative papers tend to be rejected less often for publication than non-collaborative submissions \cite{fox1991gender, presser1980collaboration}. Beyond this, collaborative research also ties in to the epistemological benefit of intersubjective verifiability and the idea that arriving at a conclusion through the unforced agreement of multiple parties further legitimizes conclusions from research \cite{beaver2004does}. Finally, yet another perceived benefit of collaborations is that they help mobilize knowledge and promote its diffusion as contributors participate in networks of knowledge exchange \cite{breschi2003mobility, ponds2009innovation}.

When examining collaborations in the scientific literature, the multiple author publication (co-author publication) is most often used as a proxy for measuring collaboration \cite{katz1997research} (e.g.  \cite{smith1958trend, de1966collaboration, clarke1964multiple} as early examples). Using this definition of collaboration has advantages, as noted by Subramanyam \cite{subramanyam1983bibliometric} and further explained by Katz \cite{katz1997research}:
\begin{enumerate}[noitemsep, topsep=0pt]
    \item it is invariant and verifiable, which makes it reproducible
    \item it is easily and inexpensively ascertainable, which it makes it practical
    \item it is quantifiable and detectable, which makes it scalabale to large scholarly data
    \item it is non-reactive, which makes the analysis thereof detached from the process being examined
\end{enumerate}
Though, as Katz notes, there may be arguments against the last of these points (e.g. \cite{persson2004inflationary}), using co-authorship as a proxy for collaboration provides a metric that is reproducible, practical, and scalable. It is of little wonder why this has become the standard definition for examining collaboration in literature, especially in eras of limited article metadata.

That said, while examining collaborations through the lens of co-authorship has become a standard of sorts, there are numerous alternate definitions that are yet unexplored. These include examining collaborations across the numerous scales at which researchers collaborate, including across/within organizations, regions, and disciplines \cite{frenken2005citation}. Much of this is now possible with increasingly better curated article-level metadata and large-scale bibliometric data. For example, Gazni and Didegah \cite{gazni2011investigating} examined the different types of research collaboration found within a single institution, extending well beyond just the multiple-author definition. We aim to expand this idea by including an analysis of research collaborations at a single institution in a context that we believe is of yet unexplored - that of inter-departmental collaboration.

The purpose of this work is to examine scientific collaborations at a large, public US research university both in the context of co-authorship as well as in the context of organizational/departmental structure. We do this by exploring two ideas: 1) whether the definition of ``collaboration'' impacts differences in citation metrics between collaborative and non-collaborative works and 2) whether the similarity of fields/departments (as defined by co-citations) signals any trend in co-authorship/collaboration. We begin with a large set of metadata on scientific papers and, after extensive data curation, examine collaborations as defined in three ways: across multiple authors, across multiple institutions, and across multiple departments. While there are numerous examples of analyses based on the first two definitions (e.g. \cite{schmoch2008international, adams2005scientific, narin1991scientific, hottenrott2017first}), the last remains largely unexplored. We compare the citation counts of papers within each of these categories of collaboration to examine whether defining collaborations in increasingly nuanced ways results in differences with respect to frequency of collaboration as well as citation metrics. Then, we present proof-of-concept results in which we investigate the degree to which co-citation and co-authorship relate when examining inter-departmental collaborations. By doing so, we examine the question of whether fields that cite the same work also tend to work together in a university setting. 

\section{Methods}
The methods of this work are described as follow: first, the data cleaning and labelling of papers as belonging to the three categories of collaboration is described; then, a description of how citation counts were calculated is provided; finally, the process by which we examine the relationship between co-citation and co-authorship is detailed. The data for this work comes from the Web of Science (WoS). A local copy of all WoS paper metadata resides in a MySQL database managed by the DataLab of the Information School at the University of Washington. The data includes publications dating back to 1900 and is current through 2017. There are over 62 million papers and hundreds of millions of citation links in the data.

\subsection{Data Cleaning and Classifying Papers}

As mentioned above, we center our analysis on a single university - the University of Washington in Seattle (UW). The motivation behind this is two-fold. Firstly, in extracting university subaffiliations and departments, there was a great deal of data curation/cleaning needed, as detailed below. This data cleaning required considerable effort and energy with respect to handling various edge cases around departmental affiliations, much of which relied on implicit knowledge of the UW and its history. The fact that two of the three authors have been at the UW for about a decade each as students/staff aided immensely in this data cleaning. To limit the extent of data cleaning, only papers published after the year 2008 and before 2016 were included in the analysis. The second reason for limiting the analysis to a single university was because a significant portion of this paper focuses on work that is relatively new in design. Because of this, we wanted to scope our work to a subset of data that was tractable as we presented preliminary results.

In this work, we define a paper as being collaborative in three different ways. The first of these is examining papers that have multiple authors, which is line with the more common definition of a collaborative paper as found in literature. Using this definition, papers are labelled as being either ``multi-author'' or ``single author'' papers. The second definition is examining papers that have authors representing more than one institution, one of which is the UW. Using this definition, papers are labelled as being either ``multi-institutional'' or ``single institution'' papers. The last is examining papers that have authors representing more than one department at the UW. These papers are labelled as being either ``multi-departmental'' or ``single department'' papers.

Every paper in the WoS dataset has metadata that includes each author's name and their affiliations (listed as organizations and suborganizations, the second of which are subgroups of the first), when available, as well as the paper's date of publication. WoS also contains citation information for every paper, including citing works and works that were cited. To extract the data for this paper, all institutional affiliations (i.e. organization listings) for all authors in WoS were gathered and any representation of the UW within this list of institutional affiliations was found. This also included obvious misspellings and typos. From there, any paper with at least one author with a UW institutional affiliation was gathered. After filtering for year of publication, this resulted in 69,148 unique papers. This set of papers and all associated metadata is henceforth referred to as the ``dataset.''

To label multi-author papers, the number of unique authors for every paper in the dataset was tallied and those papers with multiple authors were labelled as such. Then, papers with at least two authors and from at least two different organizations (i.e. the UW as well as at least one external organization) were identified and labelled as multi-institutional papers. It should be noted that this included a subset of papers that did not have a listed institution for each author in the paper metadata but had at least two institutions listed among all authors, one of which was UW. On the other hand, there was a subset of papers that had only UW listed among institutional affiliations for all authors as well as a smaller subset of papers that had only UW listed among institutional affiliations for authors with available metadata and unavailable/missing institutional affiliations otherwise. The first of these were labelled as single institution papers while the second was excluded from analysis because, depending on the missing values, the paper may or may not have been a multi-institutional paper. In all, 8,337 papers were excluded from this analysis (12.1\% of all papers) while 45,662 papers were multi-institutional papers (75.1\% of papers with usable institutional affiliations) and 15,149 were single institution papers (24.9\% of papers with usable institutional affiliations).

Isolating multi-departmental papers, meanwhile, required individually curating author suborganization listings to get a set of departments and schools associated with the UW. At the onset, there were 7,458 unique UW suborganizations from the papers in the dataset, as identified by string matching. This was eventually pared to just 72 unique suborganizations and departments in the following manner: 1) identifying all the different ways in which the same department was listed (e.g. the ``The Department of Biology'' being listed as ``Dept Bio,'' ``Department of Biol,'' and ``Biol Department'') and standardizing; 2) providing a consistent representation of departments across time, which included looking at departments that merged with others (e.g. the ``Zoology Department'' is now part of the ``Department of Biology'' at UW); 3) replacing the research labs that were listed as suborganizations with their respective departmental affiliations; 4) handing specific edge cases that resulted in ambiguous departmental affiliations and/or misspellings; and 5) grouping some departments within the larger school/college they are part of. In short, this involved manually editing and curating every departmental affiliation listed for every author among the isolated papers, which came at the expense of considerable time and effort. In all, curating and organizing the data alone required was an undertaking that spanned nearly a full calendar year, whereas the analysis thereafter was completed in a considerably shorter timeframe.

For most departmental affiliations, departments were labelled in accordance with the UW's departmental listings. However, there were some instances, such as the School of Medicine and the School of Music, where the school/college level of organization was used (which, at the UW, tends to be one step above departments in the University's organizational hierarchy). This was because either the school was listed as the suborganization on papers more frequently than individual departments or because the school does not traditionally have any departments below it in the University's organizational hierarchy. Multi-departmental papers were labelled as those with at least two authors representing at least two different departments whereas single department papers were labelled as those with only a single author or only a single department represented. It should be noted that the number of institutions represented by the authors was not taken into account when examining multi-departmental papers (i.e. multi-departmental papers were not a subset of multi-institutional papers but were an independent classification). Both multi-departmental and multi-institutional papers, however, had at least two listed authors.

\subsection{Calculating Citations}
As mentioned above, paper metadata from WoS contained information on every cited and every citing work for each paper in the dataset. From this citation data, a citation network was created by linking every paper included in the dataset with all citing (in-citations) and cited (out-citations) works. Citation counts for every paper in the dataset were calculated based on the number of in-citations. The out-citations, meanwhile, were used to calculate the co-citation matrix described below. When comparing citation counts, median and mean values were calculated and statistical tests were not used as the large number of samples would have resulted in even minute differences across groups to be statistically significant. Instead, we focus on the practical significance of the results. It should be noted that the citation network only included works that were cataloged by WoS.

\subsection{Co-citation and Co-authorship}
To further examine collaborations across departments, we also present preliminary results by comparing co-citation frequency and co-authorship frequency across departments. To do so, we first used the citation network to find all papers that were cited by the papers from the dataset (i.e. all out-citations of papers in the dataset). From there, the cited papers were linked to departments based on the departmental affiliations of the authors in the citing paper. In all, 1,107,065 unique cited works were found. To look at co-citation frequency, these papers were pared down to papers that were cited at least twice by the papers in the dataset and were published after the year 1990. This ultimately gave 45,150 unique works that were cited. From there, a 73-by-45,150 department-paper matrix was constructed which was then transformed to a 73-by-73 co-citation matrix, wherein the values of the matrix represented the number of times a department along the rows co-cited a work with a department along the columns (and vice versa). This matrix was then normalized by dividing by row sums. After this, a co-author matrix was produced by first creating a 73-by-69,148 department-paper matrix consisting of co-authorship occurrences for the papers in the dataset. From this, a 73-by-73 co-authorship matrix was constructed wherein the values of the matrix represented the number of times a department along the rows co-authored a work with a department along the columns (and vice versa). As with the co-citation matrix, this co-authorship matrix was normalized by row sums.

To produce Figure \ref{fig:matrix}, a dendrogram was created from the co-citation matrix with a linkage matrix based on the Ward variance minimization algorithm \cite{ward1963hierarchical}\footnote{This is an option in python's SciPy library to generate a linkage matrix}. The created dendrogram was used to cluster similar fields based on co-citations. When clustering, departments without at least 20 out-citations were excluded as we felt we did not have sufficient data to appropriately situate these departments for clustering from the co-citation matrix. The University's School of Medicine was also excluded from clustering because the number of papers associated therewith (along with co-citations and co-authorships) greatly skewed results. The absolute distances between groupings were excluded from presentation as they are not of large concern for this work - instead, we focus on the structure of the clustering/groupings themselves. A heatmap was then generated from the co-author matrix described above and its rows/columns were organized in accordance with the dendrogram. As such, clustered groupings were maintained across the rows/columns of the heatmap and the heatmap showed the proportion of papers across each row with a co-authorship in the corresponding column.

\section{Results and Discussion}
The results and discussion for this work is framed a manner similar to the Methods section: first, an overview is given on how papers were classified according to the various definitions of collaboration; then, the discussion focuses on citation counts across the different definitions of collaboration; lastly, the association between co-citation and co-authorship is discussed. 

\begin{figure}
    \includegraphics{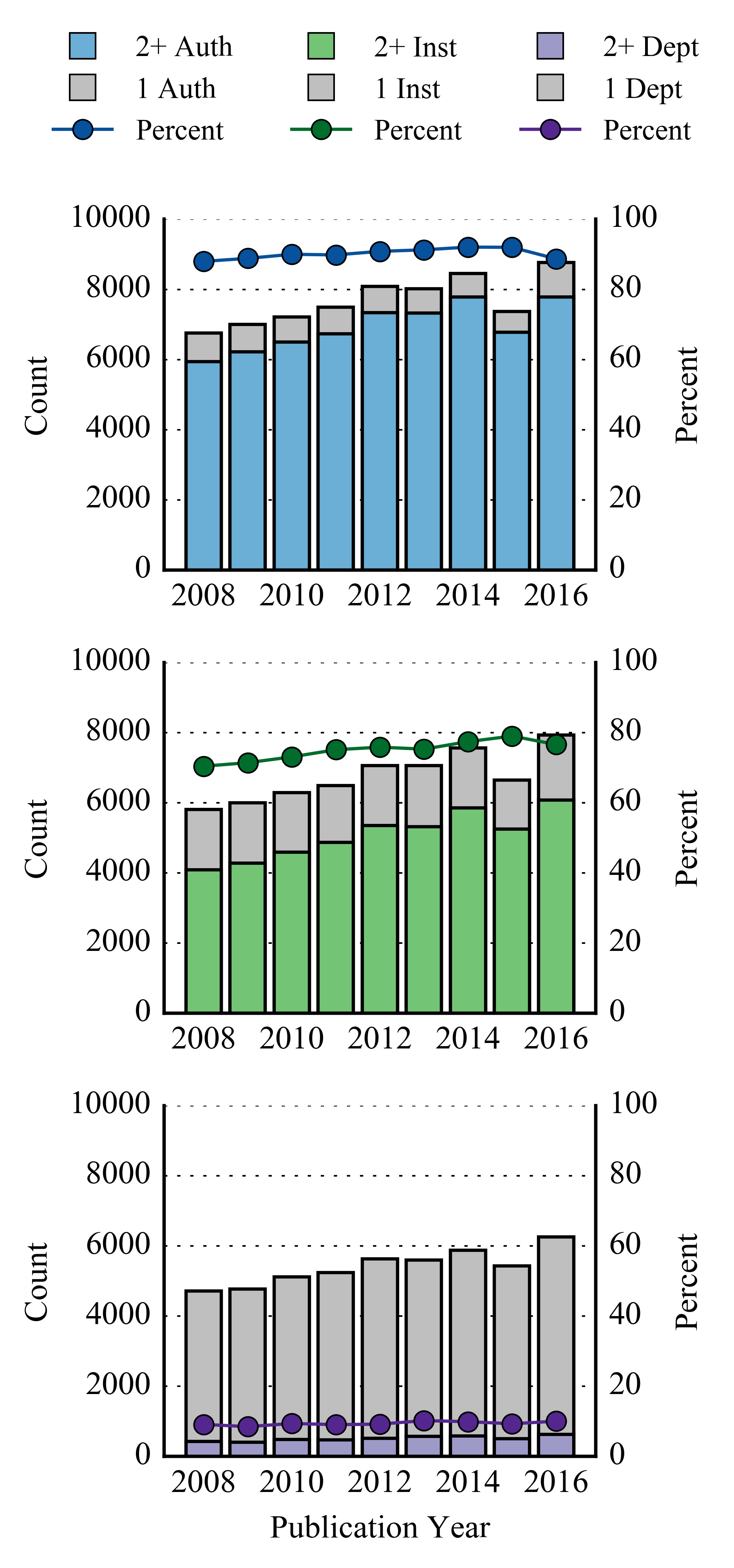}
    \caption{Counts and percentages of collaborative papers as defined by multiple authorship (top), multiple institutional affiliations (middle), and multiple departmental affiliations (bottom).}
    \label{fig:paperTime}
\end{figure}

\subsection{Classifying Papers}
The counts of all papers for each definition of collaboration is given in Table \ref{tab:paperCounts}. The counts of papers by each definition across time is shown in Figure \ref{fig:paperTime}. Of note is the relative consistency in the proportion of papers that were classified according to each definition of collaboration across time - about 80-90\% of papers by year were classified as having multiple authors, about 70-80\% of papers by year were classified as having multiple institutions represented, and about 10\% of papers by year were classified as having multiple departments. Interestingly, these consistent proportions seem to point to some level of saturation with respect to the extent of collaboration across the papers examined, which counters the notion that collaboration has been an ever-increasing research practice. When examining the total number of papers published, we are unsure of why there is a drop in total papers published for the year 2015 but do not believe that it is an artifact of our data cleaning/curation efforts.

\begin{table}
  \caption{Counts of papers by grouping}
  \label{tab:paperCounts}
  \begin{tabular}{lcc}
    \toprule
    & Count & Pct\\
    \midrule
    All Papers & 69,148 & --\\
    \hspace{0.3cm} Papers w/ 2+ authors & 62,379 & 90.2\% \\
    \hspace{0.3cm} Papers w/ 1 author & 6,769 & 9.8\% \\
    \midrule
    Papers w/ inst. affiliations & 60,811 & 87.9\% \\
    \hspace{0.3cm} Papers w/ 2+ institutions & 45,662 & 75.1\%* \\
    \hspace{0.3cm} Papers w/ 1 institution & 15,149 & 24.9\%* \\
    & & \\
    Papers w/o inst. affiliations & 8,337 & 12.1\% \\
    \midrule
    Papers w/ dept. affiliations & 48,616 & 70.3\% \\
    \hspace{0.3cm} Papers w/ 2+ departments & 4,584 & 9.4\%* \\
    \hspace{0.3cm} Papers w/ 1 department & 44,032 & 90.6\%* \\
    & & \\
    Papers w/o dept. affiliations & 20,532 & 29.7\% \\
    \bottomrule
    \footnotesize{* - percentage of subset} & & \\
\end{tabular}
\end{table}

\subsection{Citation Counts}

\begin{table}[!h]
  \caption{Citation metrics for papers by grouping}
  \label{tab:citCounts}
  \begin{tabular}{lccc}
    \toprule
    & Median & Mean & Pct w/ \\
    & citations & citations & 1+ citation \\
    \midrule
    All Papers & 5 & 19.0 & 71.7\% \\ 
    \hspace{0.3cm} Papers w/ 2+ authors & 5 & 19.9 & 73.4\% \\ 
    \hspace{0.3cm} Papers w/ 1 author & 1 & 10.3 & 56.0\% \\ 
    \midrule
    Papers w/ inst. affiliations & 5 & 19.2 & 70.8\% \\ 
    \hspace{0.3cm} Papers w/ 2+ institutions & 6 & 22.2 & 75.6\% \\ 
    \hspace{0.3cm} Papers w/ 1 institution & 1 & 10.0 & 56.1\% \\ 
    & & & \\
    Papers w/o inst. affiliations & 6 & 17.5 & 78.3\% \\
    \midrule
    Papers w/ dept. affiliations & 7 & 21.4 & 79.5\% \\
    \hspace{0.3cm} Papers w/ 2+ departments & 9 & 23.9 & 83.9\% \\
    \hspace{0.3cm} Papers w/ 1 department & 7 & 21.1 & 79.1\% \\ 
    & & & \\
    Papers w/o dept. affiliations & 1 & 13.3 & 53.0\% \\ 
  \bottomrule
\end{tabular}
\end{table}

As can be seen in Table \ref{tab:citCounts}, there is evidence that collaborative papers tended to have more citations across every level of collaboration defined. When comparing multi-author papers with single author papers, the former had an average number of citations that was nearly twice that of the latter (19.9 vs 10.3) across all papers. This difference between collaborative and non-collaborative works was even more pronounced when looking at multi-institutional vs single institution papers, wherein papers with multiple institutional affiliations had 22.2 citations on average while papers written by only those at the UW had 10.0 average citations. A similar, albeit much less pronounced, trend was seen with multi-departmental papers, wherein multi-departmental papers had an average of 23.9 citations while those with only a single department had an average of 21.1 citations. Interestingly, despite the smaller difference in citation counts between multi-departmental and single department papers, multi-departmental papers also had the highest proportion of papers with at least one citation among the groups examined at 83.9\% .

Another interesting note is that papers with more usable suborganization metadata also tended to have more citations than their counterparts. In particular, when comparing papers with usable author departmental metadata (i.e. suborganizations, as defined in WoS) with those that did not have completely usable metadata (for the purposes of disambiguating departments), the first group had an average of 21.4 citations per paper while the second had 13.3 citations per paper. This could very well be a byproduct of more highly-cited and/or visible journals/venues having better standards with respect to curating content. Interestingly, this did not hold true when comparing papers with usable institutional metadata versus those that did not have usable institutional metadata.

\begin{figure*}[!h]
    \includegraphics{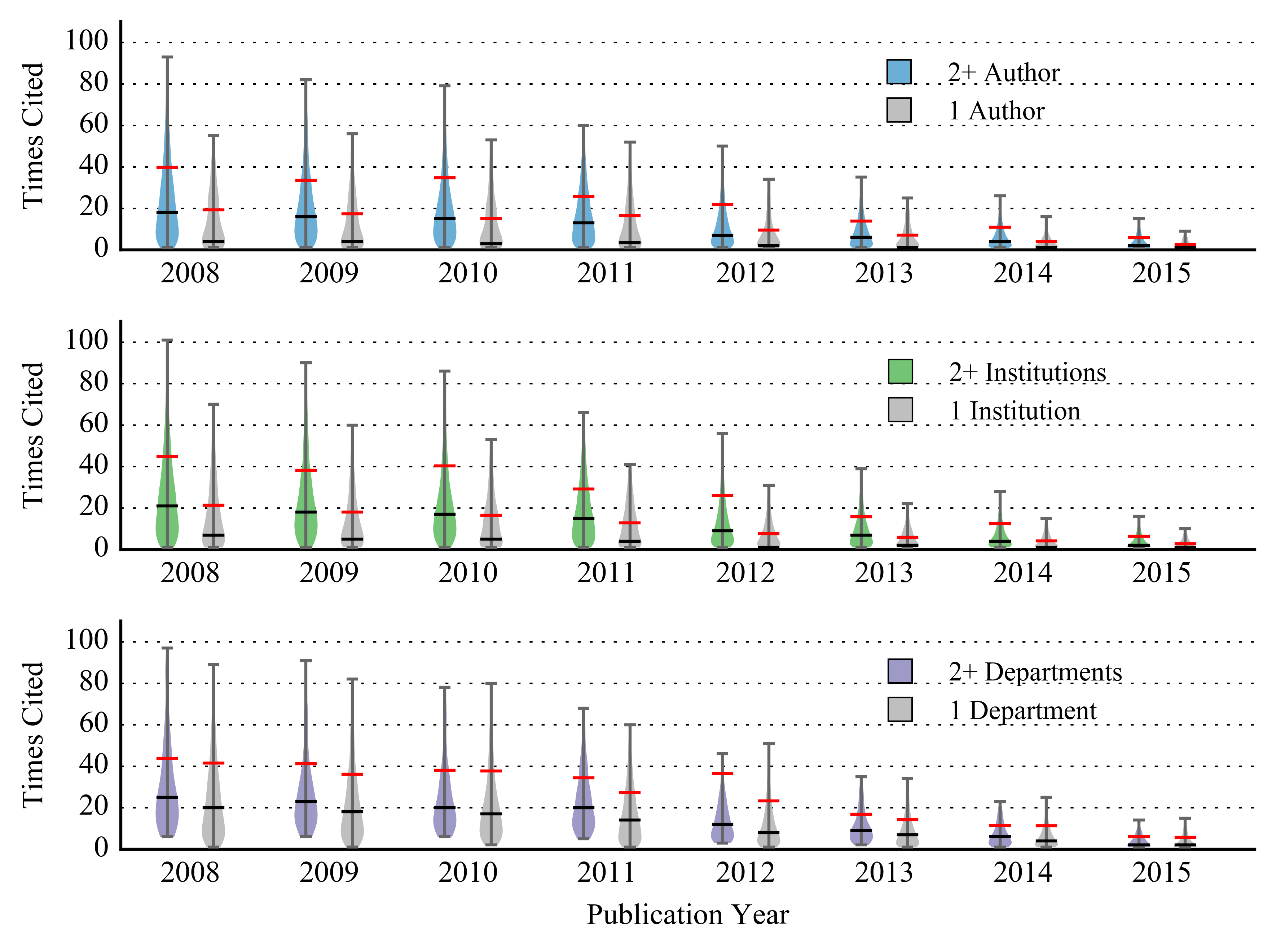}
    \caption{Violin plots of citation counts of collaborative papers as defined by multiple authorship (top), multiple institutional affiliations (middle), and multiple departmental affiliations (bottom). Horizontal black and red lines indicate median and mean values for each year, respectively. For ease of view, only values between the 10th to 90th percentiles were shown for each distribution. The medians and means were calculated using all data (i.e. not limited to values between the 10th and 90th percentiles).}
    \label{fig:citTime}
\end{figure*}

Of course, the above metrics do come with the caveat that they are not adjusted for how long the papers had to accumulate their citations. When adjusting for time of publication by only comparing papers published in the same calendar year, collaborative papers still had a greater mean and median citation count than their non-collaborative counterparts. This was true for every year examined and across every definition of collaboration. This also held true regardless of comparing the median or the mean citation counts for each group. Figure \ref{fig:citTime} shows the distributions of citations for each definition of collaboration across time, along with mean and median values. As was the case when not adjusting for time of publication, the most pronounced differences were seen when comparing multi-institutional papers with single institution papers. Comparing multi-departmental papers with single department papers again resulted in the smallest differences in citation counts between groups across time.

\subsection{Co-citation and Co-authorship}
In addition to examining citation metrics, we also wanted to understand the relationship between co-citations and co-authorships. This was mostly motivated by questions our group had with regards to multi-departmental collaborations, including which groups are reading similar literature but not collaborating at the university and whether we could predict when a collaboration is going to occur before it does. Understanding these relationships or the potential thereof can lead to more informed decisions by university and department/college administrators when allocating resources and planning around departmental research\footnote{Our research group has already begun  assisting our home institution in examining this (\url{http://www.washington.edu/global/})}. Furthermore, regions of potential cross-campus collaborations could also be future areas of new investment and growth on research campuses and outlining these areas sooner could have many long-term institutional benefits. This all also says nothing of the relationship between these departments with respect to teaching and student course-taking patterns/behavior, which we believe to be another area of great potential for research. In light of these ideas, we wanted to begin exploring whether co-citation can signal co-authorship using this data.

\begin{figure*}[!htp]
    \includegraphics[angle=270,origin=c]{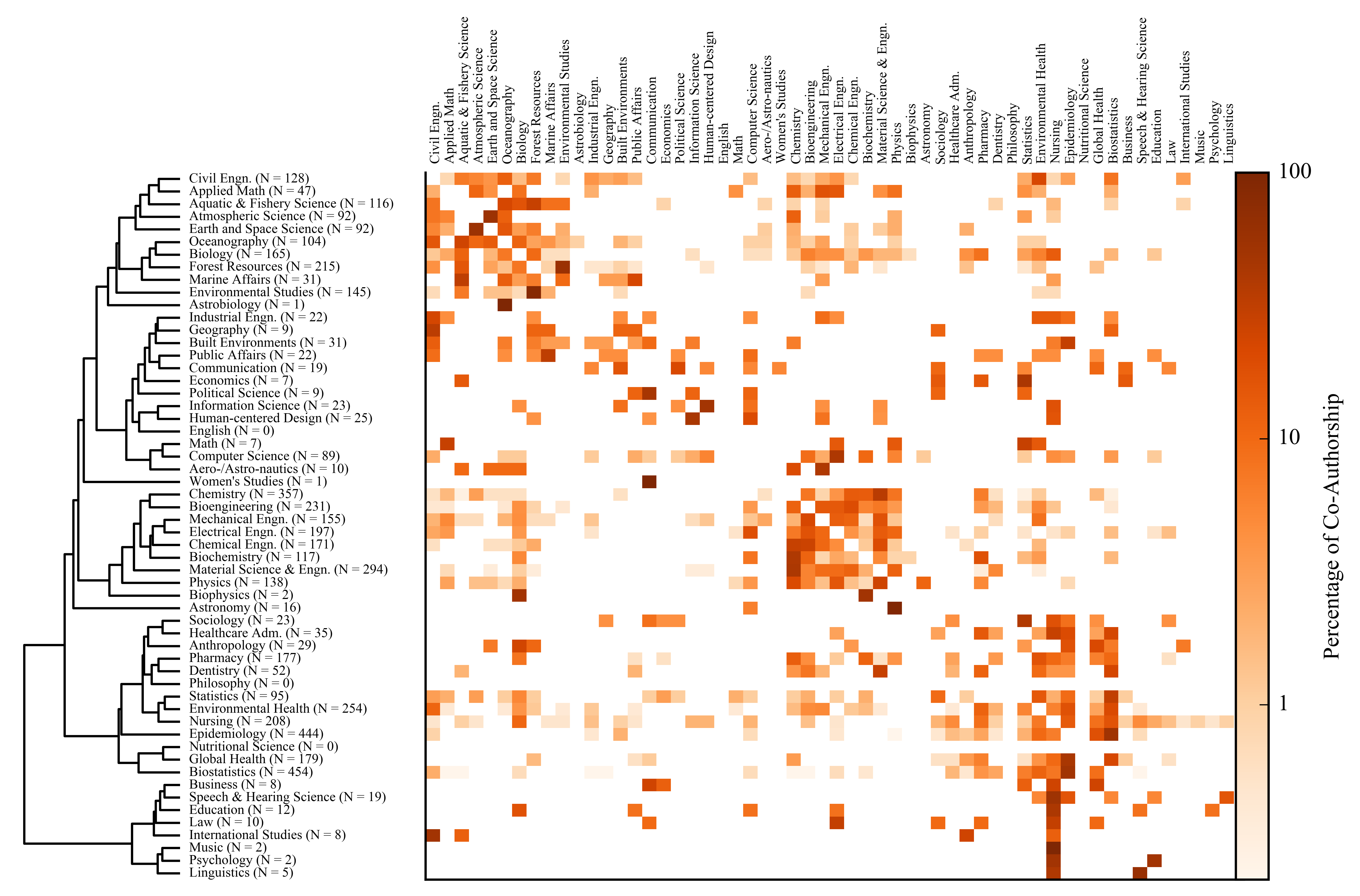}
    \caption{Dendrogram showing clustering of departments based on co-citations and heatmap showing co-authorship. Department/school titles were replaced with the fields they focus on for ease of reading. Numbers next to labels indicate number of papers co-authored with another department/school. Figure rotated 90 degrees clockwise.}
    \label{fig:matrix}
\end{figure*}

Figure \ref{fig:matrix} shows preliminary results from an exploration of the relationship between co-citation and co-authorship among departments. The dendrogram of the figure shows the groupings of departments based on co-citations with the understanding that fields that work in the same/similar domains will tend to cite the same work. As such, almost all of the groupings follow an intuitive ordering. For example, almost all of the engineering and hard sciences are clustered together. Additionally, much of the health sciences is also clustered together while biology is clustered alongside ocean, forest, and environmental sciences. Of course, there are also some peculiarities, such as the positioning of women's studies and English. We believe having more expansive data by removing limits on the data we used to develop the underlying co-citation matrix will help delineate these clusterings/groupings even further.

The figure's heatmap, meanwhile, shows the percentage of co-authorship across departments. This approach is similar to one our group has previously used when examining the underlying departmental structure of an institution of higher education with respect to students' major preferences \cite{aulck2017attrition}. There are a few things of note with respect to the heatmap. First, the diagonal of the heatmap was zeroed. Also, the heatmap was not normalized along the diagonal (i.e. the heatmap was kept asymmetric) to reflect differences in proportions across departments that have collaborated. In this sense, the heatmap shows the proportion of papers from a department along a row that had a co-authorship with a department along a column. This is not the same as the value one would get by swapping the rows/columns as the heatmap was normalized by row. Additionally, the ordering of the rows and columns of the heatmap was kept consistent with the dendrogram. That is to say that there exist distinct clusters within the row/column groupings of the heatmap based on the co-citation patterns of the departments (and, more specifically, the clustering therefrom). Furthermore, these clusters should present themselves along the diagonal of the heatmap if there is a strong relationship between co-citation and co-authorship for given departments and should be away from the diagonal otherwise. In other words, if departments are similar with respect to both co-citation and co-authorship, we should see pockets of higher values near the diagonal. 

What we see in the heatmap are extreme values along the diagonal along with some high values along the periphery of the heatmap. In particular, there appear to be three distinct clusters of co-authorship that are also reflective of co-citation: 1) among the ocean, forest, and environmental sciences; 2) among the engineering and hard sciences; and 3) among the health sciences. At the same time, there are also a few pockets of co-authorship that are away from the diagonal, namely: 1) between the ocean, forest, and environmental sciences and the engineering and hard sciences; 2) between the health sciences and the engineering and hard sciences; and 3) between the ocean, forest, and environmental sciences and the health sciences. In short, there appear to be three groups that are highly collaborative amongst themselves and are also collaborative across each other. Much of the rest of the heatmap, meanwhile, remains unfilled, indicating little in terms of collaboration. This appears to indicate siloed/compartmentalized communities within the institutional research landscape with most cross-departmental intra-institutional research spurred by three clusters of departments/fields. Our group hopes to further examine how these research relationships have changed over time with respect to clustering and compartmentalization of collaboration.

\section{Limitations}

One limitation with this work is the fact that it focuses on a single institution and the research published by it. This limits the extent to which the results may be generalized to other universities or, more broadly, the scientific literature at large. As mentioned previously, we deliberately scoped this project to focus on a single institution due to the amount of effort required to clean/curate the departmental data and in so doing, we were able to present what we believe is the first look at inter-departmental collaborations as well as the relationship between co-citations and co-authorship when examining intra-institutional collaborations. We believe this process and analysis can be expanded to other institutions to get a much broader sense of how particular fields/departments collaborate and we will soon begin examining these collaborations beyond a single institution. Additionally, we believe looking at a case study of a single university can still be beneficial to the scientific community in sparking new ideas (e.g. this work being influenced by Gazni and Didegah, who similarly looked at a single institution \cite{gazni2011investigating}).

The number of authors was also not accounted for when examining citation counts. This will likely have some impact on citation counts with respect to self-citation - if more authors are present on a work and authors are likely to self-cite in the future, the number of citations would increase accordingly. Field variation was also not accounted for in this work. This includes citation differences across fields (e.g. number of citations typically in papers, tendency to self-cite, etc.). In expanding this work, we believe controlling for self-citations and field-level differences to be an immediate area where this work can be improved.

Another consideration for this work is that lists of authors and paper metadata provide no indication of the amount of effort or research capital contributed by each individual. More exhaustive descriptions of research efforts and contributions by each author could lead to better approximations of individual efforts, as noted by Smith \cite{smith1958trend}, which can then lead to a better understanding of the collaborative nature of a work. Beyond that, however, there would still be numerous intangible aspects of research and contribution that cannot be tracked, as noted by Subramanyam \cite{subramanyam1983bibliometric}. This also says nothing of the idea of social distance between researchers resulting in a lack of accreditation \cite{katz1997research} as well as the idea of ``honorary co-authorship,'' defined as ``as the listing of the names of mentors, associates, and friends on articles, even when they have not adequately contributed for authorship'' \cite{o2009honorary}. The presented results must be taken with the understanding that every listed author and listed department were equally weighted when defining a collaboration. The number of times that a department or institution was listed was not considered as only unique occurrences of each were. This approach was taken primarily to ease analysis and future efforts may examine collaborations with respect to more rigorous weighting of effort/authorship/collaboration.

Finally, this work used WoS as a single data source. Any limitations with respect to the dataset's scope and completeness would impact the final results for this work.

\section{Future Directions}
One obvious future step with this work is to remove the bounds on the data we used in the analysis. This includes removing any restrictions with date of publications for both papers in the dataset as well as papers that were cited. As mentioned above, we would also like to expand this analysis to look at additional institutions with the understanding that this will likely require replication of the data cleaning efforts required for this work. If this data cleaning requires excess effort when scaling, we may limit the scope of our work to a single department/field (e.g. computer science).

Another potential avenue of future research is examining the degree to which papers with more curated metadata tend to have more citations. Some of the results from this work, particularly those in Table \ref{tab:citCounts} indicate that papers with curated suborganization metadata tend to have more citations than those that do not. We believe this to be a result of more visible journals/venues having better curated metadata but exploring this further would be of interest.

We would also like to examine the degree to which information technology has enabled collaborations between institutions. This includes analyzing whether the geographic distance between collaborating institutions has increased over time and whether increasingly globalized communicative environments are helping foster more globalized scientific exchanges. We would like to further understand this by geolocating each institution in the WoS database and examining how the geographic distances between collaborating institutions has changed over time.

Finally, we believe the preliminary work we presented on the relationship between co-citations and co-authorships holds great promise in terms of understanding the research and organizational structure of individual institutions. Expanding on these techniques, institutions and administrators can better understand areas of internal research collaboration. We also believe that the underlying ideas can be extended to science at large to map the broader scientific/bibliometric landscape (e.g. \cite{vilhena2014finding}). We would like to further develop and refine these analytic methods with the intention of trying to uncover whether fields that are based in similar knowledge domains are collaborating and, ultimately, whether future collaborations at the field level can be predicted or even initiated with this understanding. Though there has been some work predicting collaborations at the author level (e.g. \cite{brandao2013using, zhou2017collaborator}, we believe examining the space of field-level collaborations could be of interest.

\section{Conclusions}
In this paper, we examine collaborative scientific works through three different lenses of collaboration: multi-author, multi-institution, and multi-departmental collaborations. After extensive data curation to enable us to even define multi-departmental papers, we find that the proportion of papers belonging to each definition of collaboration remains fairly stable over the last 8 years. When looking at citation counts, papers that were defined as being collaborative were more frequently cited than their non-collaborative counterparts - a fact that held across every definition of collaboration and after accounting for year of publication. Finally, we show preliminary results as we examine the relationship between co-citations and co-authorship, finding that there is a high degree of collaboration both among clusters of similar fields and across these same clusters. We hope to further expand these methods to better understand the relationship between co-citation and co-authorship.
\begin{acks}
  The authors would like to thank Clarivate Analytics for providing the WoS data used in this research. The authors would also like to thank Jesse Chamberlin of the University of Washington Informatics program for his assistance in extracting the WoS data.
\end{acks}

\newpage
\bibliographystyle{plainnat}
\bibliography{collabs}

\end{document}